# Steering Flexural Waves by Amplitude-Shift Elastic Metasurfaces

Guangyuan Su, Yunhao Zhang, Yongquan Liu*, Tiejun Wang*

*State Key Laboratory for Strength and Vibration of Mechanical Structures, School of Aerospace Engineering, Xi'an Jiaotong University, Xi'an 710049, China*

**Abstract**

As 2D materials with subwavelength structures, elastic metasurfaces show remarkable abilities to manipulate elastic waves at will through artificial boundary conditions. However, the application prospects of current metasurfaces may be restricted by their phase-only modulating boundaries. Herein, we present the next generation of elastic metasurfaces by additionally incorporating amplitude-shift modulation. A general theory for target wave fields steered by metasurfaces is proposed by modifying the Huygens-Fresnel principle. As examples, two amplitude-shift metasurfaces concerning flexural waves in thin plates are carried out: one is to transform a cylindrical wave into a Gaussian beam by elaborating both amplitude and phase shifts, and the other one is to focus the incidence by amplitude modulations only. These examples coincide well over theoretical calculations, numerical simulations and experimental tests. This work may underlie the design of metasurfaces with complete control over guided elastic waves, and may extend to more sophisticated applications, such as analog signal processing and holographic imaging.

*Corresponding author.

E-mail: liuy2018@xjtu.edu.cn (Y. Liu), wangtj@mail.xjtu.edu.cn (T. J. Wang).

# 1. Introduction

The manipulation of elastic waves has been of special interest for a long period of time, owing to their wide applications in structural health monitoring (Croxford et al., 2007; Mitra and Gopalakrishnan, 2016), aseismic design of structures (Brule et al., 2014; Semblat and Pecker, 2009), and medical ultrasonography (White et al., 2006; Tufail et al., 2011), etc. Elastic metamaterials, composed of arrays of artificially designed subwavelength structures to generate physical parameters even not found in natural materials (Park et al., 2020), can be used to implement unconventional controls over elastic waves (Goldsberry et al., 2019; Nassar et al., 2017; Zhu et al., 2014; Liu et al., 2015; Oh et al., 2017; Chen and Huang, 2015; Chen et al., 2019; Yoo et al., 2014). The working mechanism of elastic metamaterials is mainly related to their negative effective modulus and/or mass density, which is governed by the local resonance of their microstructures. This may result in high losses and strong dispersion, and thereby disserves the propagation characteristics of elastic waves. Moreover, resulted from the complex three-dimensional structures of elastic metamaterials and their bulky sizes (Chen et al., 2016), their applications are largely hindered in practice due to fabrication difficulties with high cost.

In light of that, the elastic metasurface has been recently developed to offer a preferable possibility to manipulate elastic waves (Zhu and Semperlotti, 2016; Kim et al., 2018; Su et al., 2018). Compared with metamaterials, metasurfaces are compact in size, easier to fabricate and with smaller losses (Liu et al., 2017; Li et al., 2018), because they only consist of single- or few-layer unit cells, to shape discontinuous physical fields between different regions (Chen et al., 2016; Assouar et al., 2018). Since the generalized Snell's law proposed as the theoretical basis of phase-change metasurface (Yu et al., 2011), large number of efforts have been conducted to realize extraordinary manipulations over wave fields in electromagnetics and acoustics, including the anomalous refraction and reflection (Tang et al., 2014; Li et al., 2015; Estakhri and Alù, 2016; Pfeiffer and Grbic, 2013), energy absorption (Ma et al., 2014; Li and Assouar, 2016; Seren et al., 2014), subwavelength focusing (Chen et al., 2018; West et al., 2014), polarization conversion (Yu et al., 2012; Sun et al., 2011), and holography techniques (Zheng et al., 2015; Ni et al., 2013; Tian et al., 2017).

For elastic waves, Zhu and Semperlotti (2016) first experimentally realized the anomalous refractions of guided elastic waves in plates using elastic metasurfaces composed of locally resonant unit cells. Based on compensations on the phase discontinuities, some elastic metasurfaces have been presented to realize diverse functionalities, including the mode conversion (Kim et al., 2018), beam splitting (Su et al., 2018), source illusion (Liu et al., 2017), anomalous refractions (Lee et al., 2018; Zhang et al., 2019) and energy absorption (Cao et al., 2020). It is pointed out that, these designs focus on modulating phases of metasurfaces only.

In addition to the phase change, the modulation of amplitude not only benefits the precise control of wave fields, but also broadens the application range of metasurfaces, as has been addressed in electromagnetics and acoustics (Estakhri and Alù, 2016; Tian et al., 2017). However, once the amplitude discontinuities are introduced as a new degree of freedom, the phase-compensating approaches within the scope of geometric ray theory turn outmoded in metasurface designs. As a matter of fact, Huygens' metasurfaces have been recently proposed in electromagnetics as a new generation of metasurfaces, which can perform field transformation in a more precise way by compensating both the electric and magnetic discontinuities (Estakhri and Alù, 2016; Pfeiffer and Grbic, 2013). This inspires us to design more efficient elastic metasurfaces in this paper, termed as Huygens-type elastic metasurface, which can satisfy the boundary conditions in amplitudes and phases to transform elastic wave fields. Differing from conventional transformation approaches that require bulky artificial materials, the proposed Huygens-type elastic metasurface works with planar structures of subwavelength thickness. Moreover, the theoretical prediction of transmitted wave fields is a fundamental issue following the development of metasurfaces, especially for delicate pattern designs. Meanwhile, the generalized Snell's law is developed in ray optics (Yu et al., 2011; Born and Wolf, 2013) to omit the amplitude-related terms in nature. This problem drives us to develop the prediction theory by modifying the Huygens-Fresnel principle with compensations in phase and amplitude induced by elastic metasurfaces.

In this work, we present a design rule for the Huygens-type elastic metasurface that offers both transmitted phase- and amplitude-shift profiles to achieve desired transformation of elastic wave field. Meanwhile, the prediction theory of the transmitted wave field is developed by integrating the phase- and amplitude-shift into the Huygens-Fresnel principle, namely the generalized Huygens-Fresnel principle (GHFP). To avoid challenges in realization and stability induced by active elements, we select passive elastic metasurface to perform demonstrations. We set zigzag unit cells to construct elastic metasurfaces covering a $2\pi$ span in phase shift and 0 to 1 in amplitude modulation. Two typical examples are carried out theoretically, numerically and experimentally to demonstrate the functionalities of amplitude- and phase-shift elastic metasurfaces. One is to transform a cylindrical wave into a Gaussian beam, which seems to be unrealizable using elastic metasurfaces with only phase modulation (Zhu and Semperlotti, 2016; Kim et al., 2018; Su et al., 2018; Li et al., 2018; Liu et al., 2017; Lee et al., 2018; Cao et al., 2020). The other one is to focus flexural waves only by amplitude-shift unit cells. The focusing effect is quite robust even when we place obstacles behind the metasurface. It is noted that the reported extraordinary steering of elastic metasurfaces can be well understood by the proposed GHFP. We expect that this work underlies the designs of elastic metasurfaces with more accurate capabilities of controlling guided elastic waves, and facilitates to broaden the application prospect of elastic metasurfaces.

## 2. Theory of Huygens-type elastic metasurfaces

We consider the excitation flexural wave field $W_1 = A_1 \exp(i\varphi_1)$ in a thin plate stimulated by a source $P_0$, as shown in Fig. 1, where $i$ is imaginary unit, symbols $A_1$ and $\varphi_1$ denote the spatial distributions of the amplitude and phase in the excitation field, respectively. For simplicity, the time-harmonic dependence $\exp(-i\omega t)$ is neglected throughout this paper, where symbol $\omega$ is working angular frequency. In Fig. 1, $\boldsymbol{n}$ denotes a unit normal vector towards the transmitted field, while $\boldsymbol{r}$ and $\boldsymbol{s}$ are position vectors. Meanwhile, a target field $W_2 = A_2 \exp(i\varphi_2)$ is separated from the excitation field by a surface of arbitrary shape $\Sigma$, where symbols $A_2$ and $\varphi_2$ represent the amplitude and phase distributions in the target field, respectively. Since the excitation and target fields are discontinuous at the interface $\Sigma$, a transmission coefficient $t = A_2/A_1 \exp(i\Delta\varphi)$ at interface is needed to satisfy the boundary conditions, where phase shift $\Delta\varphi = \varphi_2 - \varphi_1$ and transmission coefficient amplitude $|t|$ represent the discontinuities in phase and amplitude, respectively. According to the Huygens principle (Born and Wolf, 2013), each point on the metasurface $\Sigma$ functions as a secondary source to constitute new wavefront. It implies that the transformation from excitation field into target field can be attained once we artificially engineer an elastic metasurface $\Sigma$ by offering the profiles of both amplitude-shift $|t|$ and phase-shift $\Delta\varphi$. Generally, the metasurface is composed of a layer of discretized subwavelength unit cells with needed local amplitude- and phase-shift. For the ease of fabrication and stability, we select passive elements to constitute the Huygens-type elastic metasurfaces in this paper. In this circumstance, the amplitude-shift profile needs to be normalized by $|t_n| = |t|/\max(|t|)$.

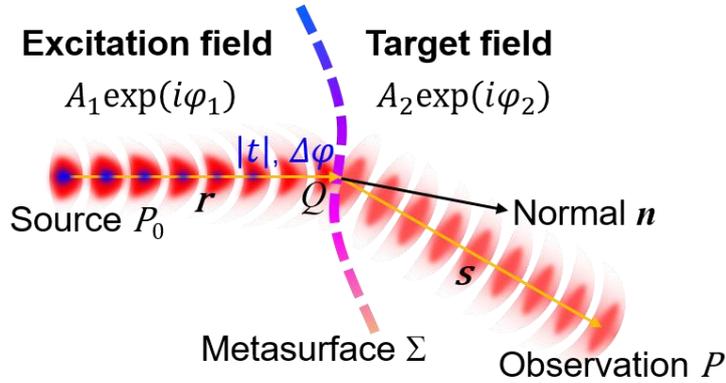

Fig. 1. Excitation and target fields separated by a Huygens-type elastic metasurface of arbitrary shape $\Sigma$.

Compared with the elastic metasurfaces designed based on the generalized Snell's law only to compensate the phase discontinuities (Zhu and Semperlotti, 2016; Kim et al., 2018; Su et al., 2018; Li et al., 2018; Liu et al., 2017; Lee et al., 2018; Cao et al., 2020), the proposed Huygens-type elastic metasurface can attain more complete and precise control over waves. It is also noted that the generalized Snell's law, developed in the scope of ray optics, is accurate only when the wavelength is much smaller than the size of structures (Born and Wolf, 2013). In this regard, replacing the generalized Snell's law with a new theory is especially necessary for elastic waves, because the wavelength of typical elastic waves (e.g., Lamb waves in plates) is usually macroscopic and large enough compared with structures (e.g., defects in plates).

To prove the design methodology, an intuitive theory for predicting the target wave pattern is in demand. The Huygens-Fresnel principle quantitatively describes the wave propagation in such a way that every point on a wavefront acts as a secondary point source, and the sum of secondary waves forms subsequent new wavefronts (Born and Wolf, 2013). As a generalization, if we view the metasurface $\Sigma$ as the elementary wavefront and further introduce the discontinuities in amplitude and phase into the Huygens-Fresnel principle, we can obtain the perturbation distribution in the transmitted field shaped by the passive Huygens-type elastic metasurface, i.e. the GHFP,

$$w = -\frac{i}{2\lambda}\int |t_n| A_1 \frac{\exp[i(\varphi_1 + \Delta\varphi + k|\boldsymbol{s}|)]}{\sqrt{|\boldsymbol{s}|}} [\cos(\boldsymbol{r},\boldsymbol{n}) + \cos(\boldsymbol{s},\boldsymbol{n})] \mathrm{d}\Sigma \quad (1)$$

where $\lambda$ is the working wavelength, $k$ the wavenumber, and the terms $A_1 e^{i\varphi_1}$, $|t_n|e^{i\Delta\varphi}$ and $e^{ik|\boldsymbol{s}|}/\sqrt{|\boldsymbol{s}|}$ account for the propagation of source to the metasurface $\Sigma$, the compensation of metasurface in phase and amplitude discontinuities, and the transmission of secondary wave to observation position $P$, respectively. The term $\cos(\boldsymbol{r},\boldsymbol{n}) + \cos(\boldsymbol{s},\boldsymbol{n})$ is an inclination factor to describe the direction dependent features of the amplitude of the secondary waves. Once the discontinuities in both amplitude and phase are vanished by letting $|t_n| = 1$ and $\Delta\varphi = 0$, the GHFP is degenerated to the

classical Huygens-Fresnel principle.

Compared with previous theory of Huygens' metasurfaces which stipulate electric and magnetic impedances to control electromagnetic fields (Estakhri and Alù, 2016; Pfeiffer and Grbic, 2013), Eq. (1) refers to a single transmission coefficient $t_n$ modulating wave propagation behaviors. This reduced version may be easy-to-use for the field transformation and thus can be easily generalized to acoustic counterparts. Particularly, if we eliminate the amplitude-shift effect as the reduced GHFP with unitary transmission coefficient, the generalized Snell's law can be comprehended in a more fundamental form of wave propagation (see details in Appendix A).

## 3. Transformation of a cylindrical wave into a Gaussian beam

Previous elastic metasurfaces are restricted to shape wave-fronts by compensating phase discontinuities (Zhu and Semperlotti, 2016; Kim et al., 2018; Su et al., 2018; Lee et al., 2018; Zhang et al., 2019), for which the theoretical basis is the geometric ray theory. In this section, we design a Huygens-type elastic metasurface transforming a cylindrical wave into an oblique Gaussian beam, which has been addressed in acoustics as a typical demonstration for the amplitude modulation (Tian et al., 2017). The transmitted wave fields calculated by GHFP and full wave simulations are presented to check the validity. The underlying mechanism of selected unit cells fulfilling needed responses in phase and amplitude is revealed as well.

*3.1. Metasurface design and theoretical prediction*

As shown in Fig. 2(a), a cylindrical wave excited by a point source $P_0$ of unitary amplitude impinges an elastic metasurface placed along $y$ axis. An oblique Gaussian beam centered at origin $O$ is supposed to be generated. The spatial distributions of amplitude and phase in incident field are $A_1 = 1/\sqrt{|r|}$ and $\varphi_1 = k|r|$, respectively, where $|r| = \sqrt{(x-x_0)^2 + (y-y_0)^2}$ with $(x_0, y_0)$ being the position of point source $P_0$. Here, the point source $P_0$ is exerted at (-150, 0) mm. The expressions for the amplitude and phase distributions in transmitted field are (Serdyuk and Titovitsky, 2010)

$$A_2(x,y) = \exp\left[-\frac{1}{\omega_G^2}(y - \tan\theta_G x)^2\right]$$
$$\varphi_2(x,y) = k(\cos\theta_G x + \sin\theta_G y) \tag{2}$$

where $\theta_G$ is the oblique angle and $\omega_G = \omega_0/\cos\theta_G$ is the Gaussian beam half-width in the $y$ direction. According to the discontinuities on the interface between incident and transmitted fields, we obtain the profiles of phase shift and transmission coefficient amplitude

$$\Delta\varphi(y) = k\left[\sin\theta_G\, y - \sqrt{x_0^2 + (y-y_0)^2}\right]$$
$$|t_n(y)| = \frac{\sqrt[2]{x_0^2 + (y-y_0)^2}}{\sqrt{|x_0|}} \exp\left(-\frac{y^2}{\omega_G^2}\right). \tag{3}$$

According to GHFP, the transmitted out-of-plane displacement field is theoretically predicted by

$$w = -\frac{i}{2\lambda}\int_{B_l}^{B_u} |t_n(y_m)| A_1(y_m) \frac{\exp[i[\varphi_1(y_m) + \Delta\varphi(y_m) + k|s|]]}{\sqrt{|s|}}$$
$$\times [\cos(\boldsymbol{r}, \boldsymbol{n}) + \cos(\boldsymbol{s}, \boldsymbol{n})] \mathrm{d}y_m \tag{4}$$

where $|s| = \sqrt{x^2 + (y-y_m)^2}$, $B_u$ and $B_l$ represent the upper and lower bounds for integral, respectively.

As a benchmark, we set the oblique angle $\theta_G$ is taken to be 30 degree, and the Gaussian beam half-width $\omega_0$ is $3\lambda$. Fig. 2(b) shows the target displacement field of the Gaussian beam. Throughout this paper, we stimulate the asymmetric Lamb wave ($A_0$ mode) in a 304-steel plate with thickness of 1.5 mm, and the working frequency $f$ is set to 15 kHz with the wavelength $\lambda$ measured to be 30.17 mm. Figs. 2(c) and 2(d) display the profiles of transmission coefficient amplitude $|t_n|$ and dimensionless phase shift $\Delta\varphi/\pi$, respectively. From Fig. 2(c), the distribution of transmission coefficient amplitude $|t_n|$ is symmetric about $y = 0$ (indicating $B_u = -B_l$) and decreases to near zero at $y = \pm 10\lambda$ ($|t_n(10\lambda)| = 0.0004$).

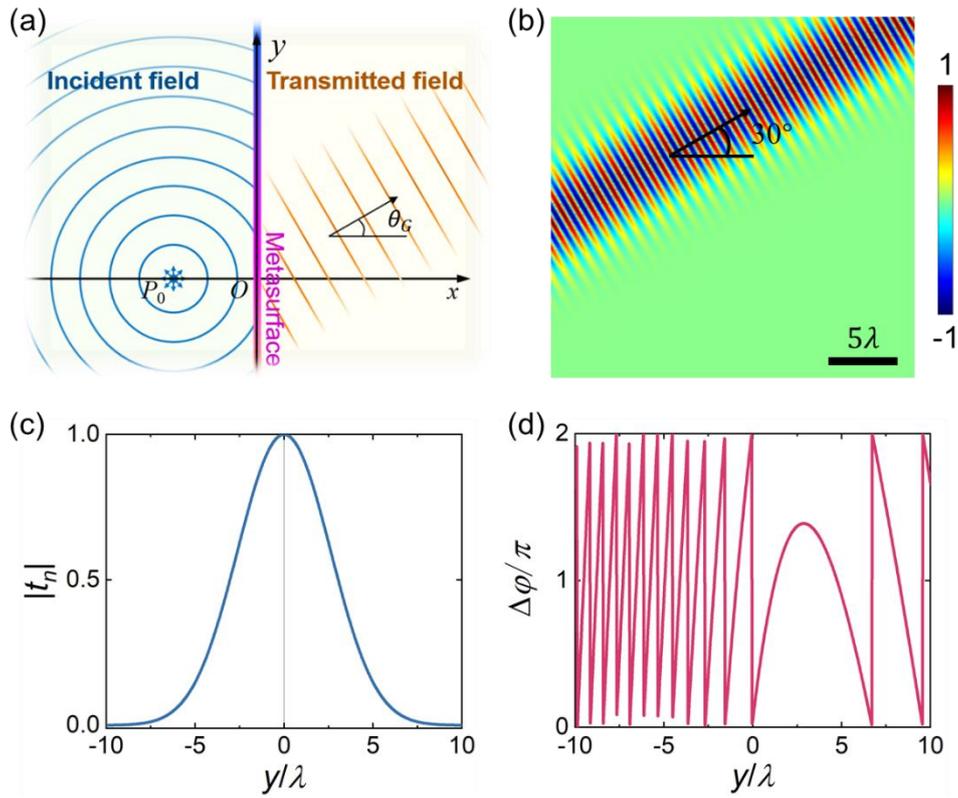

Fig. 2. Huygens-type elastic metasurface which transforms a cylindrical wave into an oblique Gaussian beam. (a) Illustration of a point source $P_0$ and a generated Gaussian beam in Cartesian coordinate system. (b) Target displacement field of the Gaussian beam. (c) Distribution of the transmission amplitude on the Huygens-type metasurface. (d) Distribution of the phase shift on the metasurface. In (b)-(d), the half-width of the Gaussian beam is $3\lambda$, the point source is exerted at (-150, 0) mm, and the wavelength λ=30.17 mm.

Fig. 3(a) shows the transmitted displacement field calculated from GHFP with integral bound $B_u = 10\lambda$, which is almost identical to Fig. 2(e). To manifest the effect of amplitude modulations, Fig. 3(b) displays the displacement field calculated with the same parameters in Fig. 3(a) except for a unitary transmission amplitude $|t_n| = 1$. One can see inconsistent amplitude distributions in wave-fronts, which indicates the necessity of amplitude modulation. Moreover, Figs. 3(c) and 3(d) show the displacement distributions with $B_u = 6\lambda$ ($|t_n(6\lambda)| = 0.06$) and $B_u = 3\lambda$ ($|t_n(3\lambda)| = 0.51$), respectively. A Gaussian beam, which is almost the same as the one in Fig. 2(b), is seen in Fig. 3(c), while the transmitted field in Fig. 3(d) is heavily influenced by the excessive truncation. This indicates that the integral bounds truncation $B_{u/l} = \pm 6\lambda$ is reasonable. In the ensuing subsections, we construct the Huygens-type elastic metasurface with unit cells of zigzag structures, and further validate the proposed theory by numerical simulations.

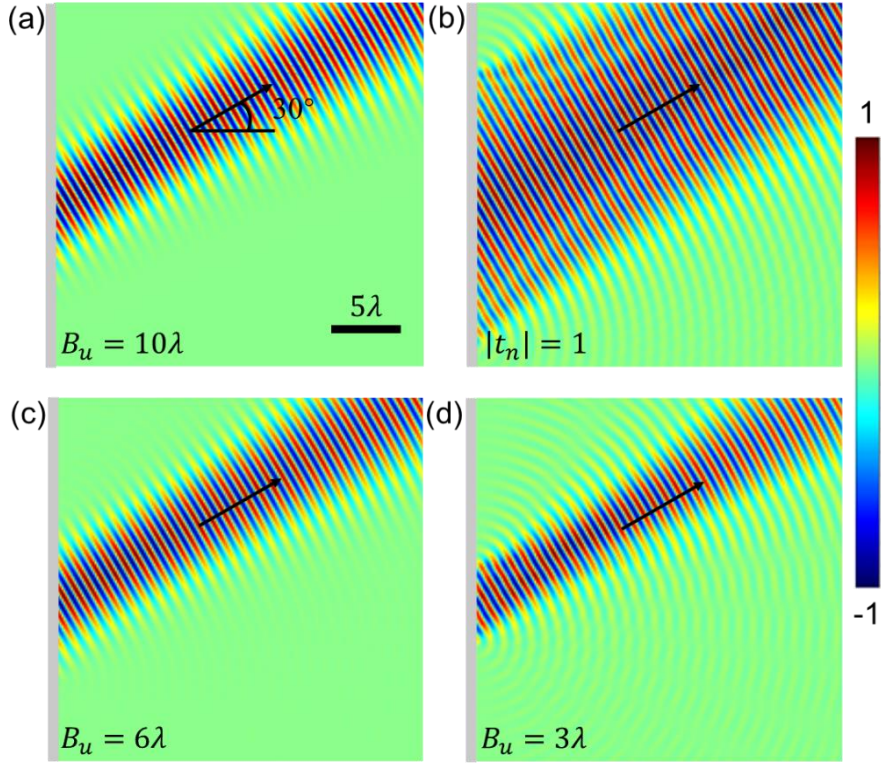

Fig. 3. Transmitted fields predicted by GHFP for the Huygens-type metasurface to generate an oblique Gaussian beam. (a) Displacement field with integral bound $B_u = 10\lambda$. (b) Displacement field with $B_u = 10\lambda$ and $|t_n| = 1$. (c) Displacement field with $B_u = 6\lambda$. (d) Displacement field with truncated bound $B_u = 3\lambda$.

*3.2.    Modulation mechanism of zigzag unit cells*

As shown in Figs. 2(c) and 2(d), the unit cells of a Huygens-type elastic metasurface need to be able to offer transmissions ranging from 0 to 1 and phase shifts covering a $2\pi$ span. We select zigzag unit cells to constitute metasurfaces, which have been confirmed of providing the complete phase shifts with high and low transmissions (Liu et al., 2017). In this subsection, we reveal in detail the tunable properties of transmitted coefficient amplitudes for zigzag unit cells.

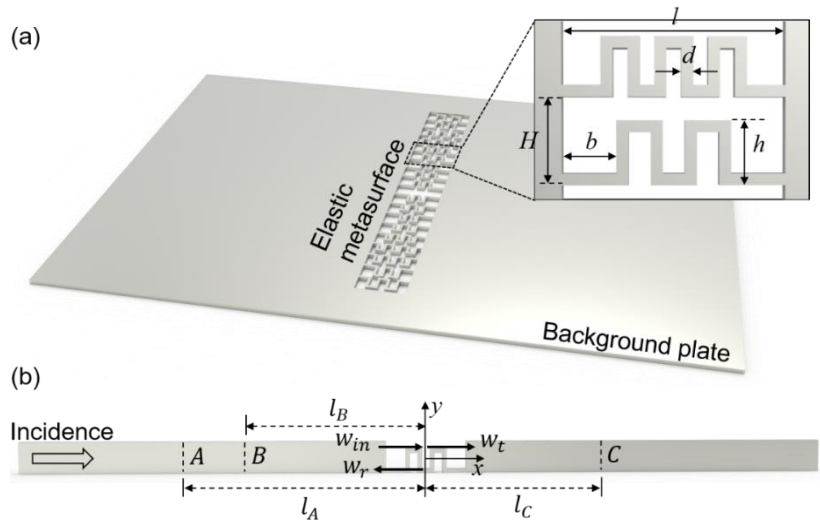

Fig. 4. An elastic metasurface composed of zigzag unit cells. (a) Illustration of an elastic metasurface consisting of two- and three-turn zigzag unit cells and the detailed geometries of unit cells. (b) Schematic diagram for derivations of the transmission and reflection coefficients of a zigzag unit cell.

Fig. 4(a) sketches an elastic metasurface composed of two- and three-turn zigzag unit cells, which is centrally embedded in a plate, and the inset displays the detailed geometries of zigzag unit cells. In order to ensure the discretization, the height of unit cells $H$ needs to be at subwavelength scale. In a unit cell, the zigzag waveguide molds the path that the wave will propagate along, ultimately resulting in the phase delay. One can adjust the height of turns $h$ and the number of turns, which lead to changes in the length of wave propagating path, to realize the needed phase shifts.

It is known that a purely theoretical analysis on the transmitted responses of a zigzag unit cell is complicated. So, a simulation-aided method is presented here. Consider a strip-like model with an embedded unit cell, as shown in Fig. 4(b). Periodic boundary conditions are applied to the top and bottom surfaces of the model, and perfectly matched layers are implemented at two ends to negate boundary reflections. A normal plane source with out-of-plane displacement is exerted at the left side of the model to excite $A_0$ mode Lamb wave. It is pointed out that the $A_0$ mode Lamb wave propagates in a zigzag unit cell without mode conversion under present working frequency, due to the symmetrical properties of zigzag structures (Liu et al., 2017). Apparently, the out-of-plane displacements at positions $A$, $B$ and $C$ can be expressed as

$$w_A = w_{in}e^{-ikl_A} + w_r e^{ikl_A}$$
$$w_B = w_{in}e^{-ikl_B} + w_r e^{ikl_B} \tag{5}$$
$$w_C = w_t e^{ikl_C}.$$

where symbols $w_{in}, w_r$ and $w_t$ denote the incident, reflected and transmitted waves of the unit cell, respectively. Once the displacements at positions $A$, $B$ and $C$ are extracted from simulations, we can easily determine the incident, reflected and transmitted waves of the unit cell

$$w_{in} = \frac{w_A e^{ikl_A} - w_B e^{ik(2l_A - l_B)}}{1 - e^{i2k(l_A - l_B)}}$$
$$w_r = \frac{w_B e^{-ikl_B} - w_A e^{ik(l_A - 2l_B)}}{1 - e^{i2k(l_A - l_B)}} \tag{6}$$
$$w_t = w_C e^{-ikl_C}.$$

The transmission and reflection coefficients are obtained by $t_n = w_t/w_{in}$ and $r_n = w_r/w_{in}$, respectively, from which we can readily determine the corresponding phase shifts $\Delta\varphi_t = \arg(t_n)$ and $\Delta\varphi_r = \arg(r_n)$.

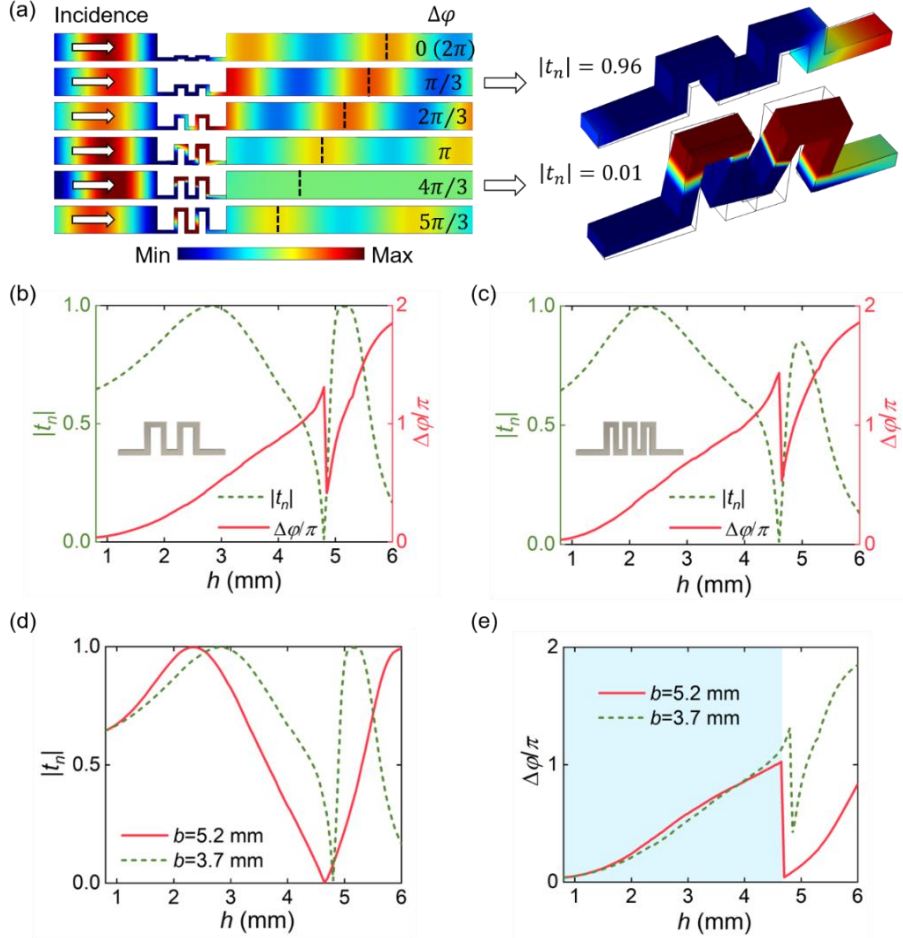

Fig. 5. (a) Simulated out-of-plane displacement fields for two-turn zigzag unit cells with $b = 3.7$ mm and varying heights $h$, and the detailed deformations of two unit cells with extreme transmissions under an identical scale factor. (b) Variations of transmission coefficient amplitudes $|t_n|$ and dimensionless phase shifts $\Delta\varphi/\pi$ with heights $h$ for the unit cells in (a). (c) Variations of transmitted amplitudes $|t_n|$ and phase shifts $\Delta\varphi/\pi$ with heights $h$ for three-turn zigzag unit cells with $b = 3.9$ mm. (d) Variations of transmitted amplitudes $|t_n|$ with heights $h$ for two-turn unit cells of two spacings $b$. (e) Variations of phase shifts $\Delta\varphi/\pi$ with heights $h$ for the unit cells in (d). Geometry parameters $d = 0.8$ mm, $l = 15$ mm ($\sim 0.5\lambda$) and $H = 6$ mm ($\sim 0.2\lambda$) are used here.

Herein, we use commercial software COMSOL Multiphysics to implement full wave simulations. The material properties used here are Young's modulus $E = 200$ GPa, density $\rho = 7900$ kg/m$^3$ and Poisson's ratio $\nu = 0.3$ for 304-steel. To manifest the modulation capacity of zigzag unit cells, Fig. 5(a) plots the simulated out-of-plane displacement fields with varying heights $h$ for the strip-like models of two-turn zigzag unit cells. It is seen that the transmitted phase shifts increase with the wave propagating path inside a unit cell, to satisfy a $2\pi$ span coverage. Different from the phase shift plots with high transmissions in the literature (Kim et al., 2018; Li et al., 2018), a significant decrease in transmission coefficient amplitude, nearly from 1 to 0, can be seen in Fig. 5(a), which suggests a transmission dip induced by resonances of the zigzag unit cells. Detailed deformations of two unit cells with extreme transmissions are displayed in the right part of Fig. 5(a), with black wireframes being the undeformed configurations. The drastic out-of-plane deformation of the unit cell with near-zero transmission reveals its twisting resonance nature.

Figures 5(b) and 5(c) display the simulated variations of transmission coefficient amplitudes $|t_n|$ and dimensionless phase shifts $\Delta\varphi/\pi$ with heights $h$ for two- and three-turn zigzag unit cells, respectively. For the two- and three-turn zigzag unit cells, the phase shifts increase with heights $h$ to almost cover a $2\pi$ span, and the transmitted amplitudes range from 0 to 1, proving the power of zigzag unit cells in amplitude and phase modulations. Fig. 5(d) displays the transmission coefficient amplitudes $|t_n|$ as functions of heights $h$ for two-turn zigzag unit cells with two different spacings $b$. The widths of transmission dips increase with spacings $b$, which is consistent with Liu et al. (2017). Besides, as shown in Fig. 5(e), the phase shifts before resonances (curves in blue area) are almost identical for the two spacings, since the lengths of propagation paths inside unit cells are identical. This suggests that the needed low transmissions can be found by adjusting spacings $b$ before resonance occurs, when the phase shifts remain unchanged for a fixed height $h$.

## 3.3. Transmitted Gaussian beam from full-wave simulation

Herein, we employ zigzag unit cells satisfying desired local responses to compose the Huygens-type elastic metasurface transforming a cylindrical wave into an oblique Gaussian beam, to further validate the theoretically calculated displacement field.

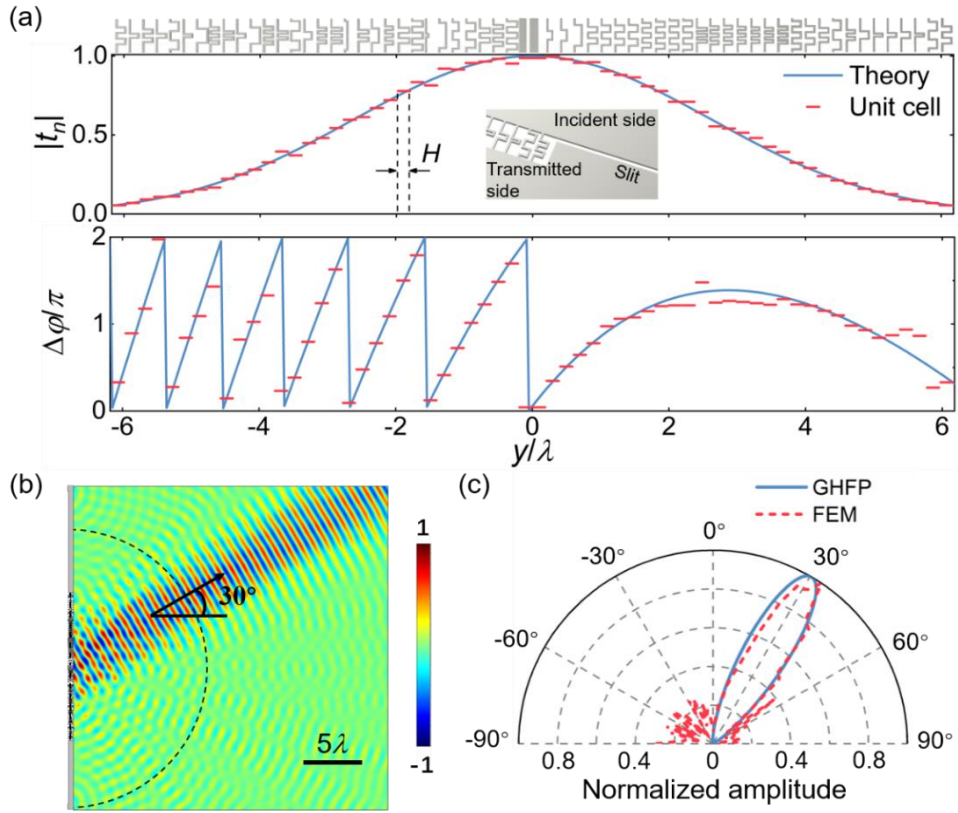

Fig. 6. (a) Theoretical and discretized profiles of transmission coefficient amplitudes $|t_n|$ and phase shifts $\Delta\varphi/\pi$ for the Huygens-type metasurface generating a Gaussian beam, with the sketches of used zigzag unit cells placed at the bottom. Inset is the setup of slit. (b) Simulated out-of-plane displacement distributions of transmitted field. (c) Polar plot of the normalized amplitudes from GHFP and simulations along the semi-circular arc marked in (b).

The transmission coefficient amplitudes $|t_n|$ and dimensionless phase shifts $\Delta\varphi/\pi$ from theory and used unit cells are shown in Fig. 6(a) as functions of dimensionless spatial coordinate $y/\lambda$, and the sketches of used unit cells with $l\sim0.5\lambda$ and $H\sim0.2\lambda$ are displayed at the bottom. The channel widths $d$ of these unit cells are 0.8 mm, except that 3 unit cells of $d = 1$ mm are marked by triangle symbols. For simplicity, the thick beam unit cells are mounted where nearly full transmission is in demand, and the zigzag unit cells with low transmissions of $|t_n| \leq 0.05$ ($|y| > 6.2\lambda$) are replaced by a slit with width of 2 mm against incident side, as shown by the inset in Fig. 6(a). Fig. 6(b) displays the simulated transmitted out-of-plane displacement field, from which one can see a Gaussian beam with an oblique angle of 30 degree. For quantitative comparison, Fig. 6(c) shows the normalized amplitudes from GHFP and simulations, along the semi-circular arc marked in Fig. 6(b). The theoretical and simulated amplitudes are normalized by their respective maximums. The results at the azimuth of 30 degree coincide well. As shown in Figs. 6(b) and 6(c), some side lobes exist, which is due to the errors in discretization.

## 4. Flexural wave focusing by only amplitude-shift modulation

The metasurfaces have shown applications in the focusing of elastic waves by acting as phase masks to shape hyperbolic phase profiles (Zhu and Semperlotti, 2016; Su et al., 2018; Li et al., 2018; Lee et al., 2018; Cao et al., 2020), which is within the scope of geometric ray theory as well. In fact, from the point view of wave motion, the amplitude distributions on metasurfaces should be engineered to realize elaborate focusing patterns, e.g., multi-focal focusing (Tian et al., 2017) and sub-diffraction near-field focusing (Chen et al., 2018). In this section, we present a flat focal lens based on the Huygens-type elastic metasurface with only amplitude shift to demonstrate the diversified potential of amplitude modulations in elastic wave focusing. The present flexural wave focusing shows robustness against obstacles arranged right behind the metasurfaces.

## 4.1. Metasurface design and theoretical prediction

The incident wave is focused by the confluence of two transmitted symmetrical Airy-like beams (Siviloglou et al., 2007). The Airy beam is first reported as a non-spreading wave packet solution of Schrödinger equation, and has been exploited in electromagnetics and acoustics because of its diffraction-free, self-healing and freely-accelerating properties (Siviloglou et al., 2007; Chen et al., 2019). The so-called freely-accelerating property means that the Airy beam lacks parity symmetry and accelerates towards one side during propagation, based on which the confluence is realized. The ideal Airy beam, composed of one main lobe and infinite side lobes, is experimentally unrealizable. One feasible way to realize Airy beam is to restrain its infinite tails by introducing an exponentially truncation function to the initial condition

$$\phi_0(y) = \text{Ai}[s(y)]\exp[as(y)] \qquad (7)$$

where $\text{Ai}[s(y)] = 1/\pi \int_0^\infty \cos[t^3/3 + ts(y)]dt$ is the Airy function, $s(y) = \beta y$ the dimensionless transverse coordinate, and $a$ the truncation parameter. The truncated Airy beam is governed by

$$\phi(x,y) = \text{Ai}[s(y) - \xi(x)^2/4 + ia\xi(x)]\exp[as(y) - a\xi(x)^2/2 \\ -i\xi(x)^3/12 + ia^2\xi(x)/2 + is(y)\xi(x)/2] \qquad (8)$$

where $\xi(x) = \beta^2 x/k$ is normalized propagation distance. The symmetrical Airy beam with respect to $y = y_s$ is easily obtained by $\phi_s(x,y) = \phi(x, 2y_s - y)$, and the interference of two symmetric Airy beams centered at $y = 0$ is $\phi_{sp}(x,y) = \phi(x, y + y_s) + \phi(x, y_s - y)$.

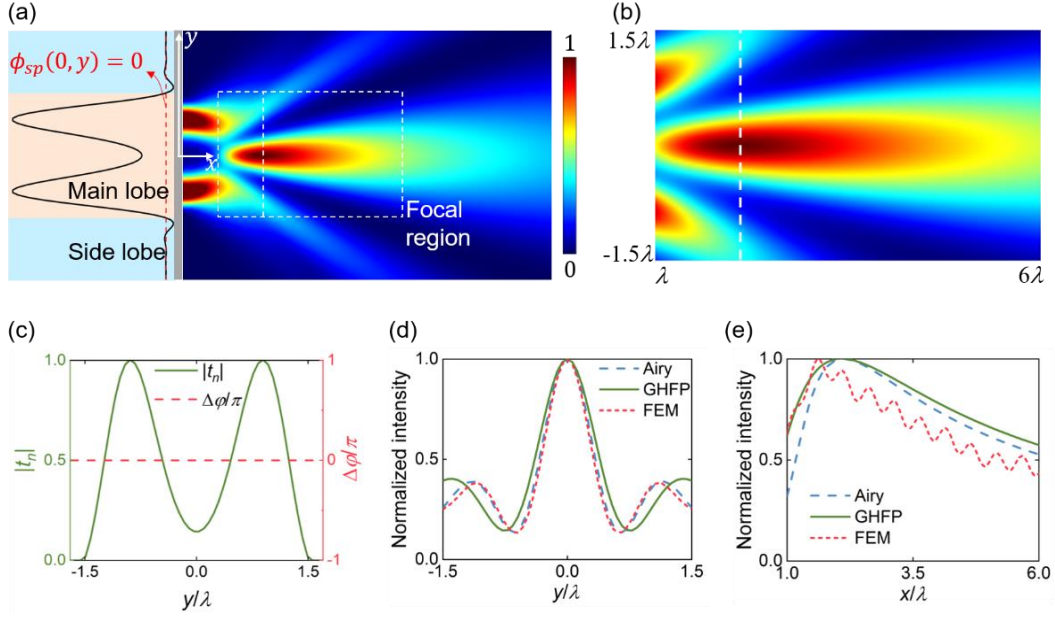

Fig. 7. Flexural wave focusing by the interference of two symmetrical Airy-like beams. (a) Normalized intensity distributions of the interference of two symmetrical Airy-like beams $|\phi_{sp}(x,y)|^2/\max\left[|\phi_{sp}(x,y)|^2\right]$ for $a = 1$, $y_s = \lambda$ and $\beta = 0.15 \text{ mm}^{-1}$. The left part displays the displacement profile at origin $\phi_0(y + y_s) + \phi_0(y_s - y)$, with orange and blue areas denoting the main and side lobes, respectively. The white dashed line in focal region denotes the cross section on focal spot in $y$ direction. (b) Normalized intensity distributions of the focal region predicted by GHFP. (c) Transmission and phase shift distributions on elastic metasurface. (d) Normalized intensity profiles along the white dashed lines in (a) and (b). (e) Normalized intensity distributions along $x$ axis in (a) and (b).

For the case of $a \ll 1$, the restrained Airy beam resembles a non-diffracted one along the propagating path (Siviloglou et al., 2007). Using phase-shift acoustic metasurfaces, Chen et al. (2019) designed a planar focusing lens by interfering two symmetric restrained Airy beams, but the focal length was more than 15 wavelengths away from the metasurface. Considering that a smaller focal length and a larger numerical aperture are crucial in application scenarios like acoustic imaging (Jin et al., 2019), we proposed a lens which can focus wave at only around $2\lambda$ behind the metasurface. We take $a = 1$ to truncate the Airy beam such that the side lobes become negligible even at a small distance to ensure a quality focal region. Fig. 7(a) displays the normalized intensity distributions of superposition $|\phi_{sp}(x,y)|^2/\max[|\phi_{sp}(x,y)|^2]$, in which one can see a focusing area marked by a white dashed box. The displacement profile at origin ($x=0$) $\phi_0(y+y_s) + \phi_0(y_s-y)$ is shown at the left side of Fig. 7(a) as well. The side lobes (blue area) are negligible in comparison with the main lobes (orange area). Since the phase of the main lobe of Airy beam keeps zero at origin, the target amplitude profile at $x=0$ with side lobes eliminated is $A_2(y) = [\phi_0(y+y_s) + \phi_0(y_s-y)]H(y+y_s - y_{lz})H(y_{rz} - y - y_s)$, where $H(\cdot)$ is the Heaviside function, $y_{lz}$ and $y_{rz}$ represent the left and right zeros for main lobes in $\phi_0(y+y_s) + \phi_0(y_s-y) = 0$, respectively.

The incident flexural wave source is a normal Gaussian beam with $\omega_0 = 3\lambda$, and the amplitude and phase profiles at $x=0$ are $A_1(y) = \exp(-y^2/\omega_0^2)$ and $\varphi_1(y) = 0$, respectively. The phase shifts of the metasurface are apparently zero, implying no phase modulations, and the transmission coefficient amplitude profile is determined by $|t_n| = A_2/A_1/\max[A_2/A_1]$. Fig. 7(c) plots the profiles of amplitude and phase shifts on metasurface. The double-peak profile of amplitudes inevitably calls for an amplitude modulation of the metasurface. The transmissions of the metasurface reduce to 0 at about $y = \pm 1.6\lambda$, indicating a truncation to the integral bound in GHFP. According to Eq. (1), we can obtain the theoretical displacement distributions in transmitted field. Fig. 7(b) shows the intensity distributions of focal region predicted by GHFP, which is coincident with the focal region in Fig. 7(a). For more detailed comparison, Figs. 7(d) and 7(e) plot the normalized intensity profiles from Airy solution and GHFP, on cross sections of the focal spots along $y$ and $x$ directions, respectively. From Fig. 7(e), the focal lengths are identical (2.1$\lambda$) for Airy solution and GHFP. The full width at half maximum of GHFP ($0.78\lambda$) is larger than that of Airy solution ($0.67\lambda$), as shown in Fig. 7(d). This is due to that the main lobes of reduced Airy beams in GHFP, which lack the shaping from side lodes, are more spreading than those in exact Airy solution.

*4.2.    Flexural wave focusing from simulations*

Fig. 8(a) plots the discretized and theoretical profiles of transmission coefficient amplitudes and phase shifts. Sketches of used unit cells are also displayed in the graph of phase shifts. For the position with near-zero transmissions ($|y| > 1.6\lambda$), the zigzag unit cells are replaced by a thin slit with width of 2 mm. Owing to the symmetry of response profiles, the unit cells are symmetrical with respect to $y = 0$. In this case, the used geometric parameters of zigzag unit cells are $l = 15$ mm ($\sim 0.5\lambda$), $H = 6$ mm ($\sim 0.2\lambda$) and $d = 0.8$ mm. For simplicity, the thick beam unit cells with width of 3.5 mm are placed where nearly full transmission is desired.

Fig. 8(b) plots the simulated normalized intensity distributions of the focal area, which coincides with Fig. 7(b). The ripple-like variation in $x$ direction is induced by the discretization over amplitude- and phase-shift profiles. The quantitative comparisons between the intensity profiles from FEM and GHFP on the cross sections of focal spots along $y$ and $x$ directions are shown in Figs. 7(d) and 7(e), respectively. The shapes and trends of intensity profiles in the two directions agree well for FEM and theoretical calculations. As shown in Fig. 7(e), the focal length in FEM ($1.6\lambda$) is smaller than that in theoretical calculations, and the intensity peak of FEM decays faster in $x$ direction than that of GHFP, which result from the errors in discretization.

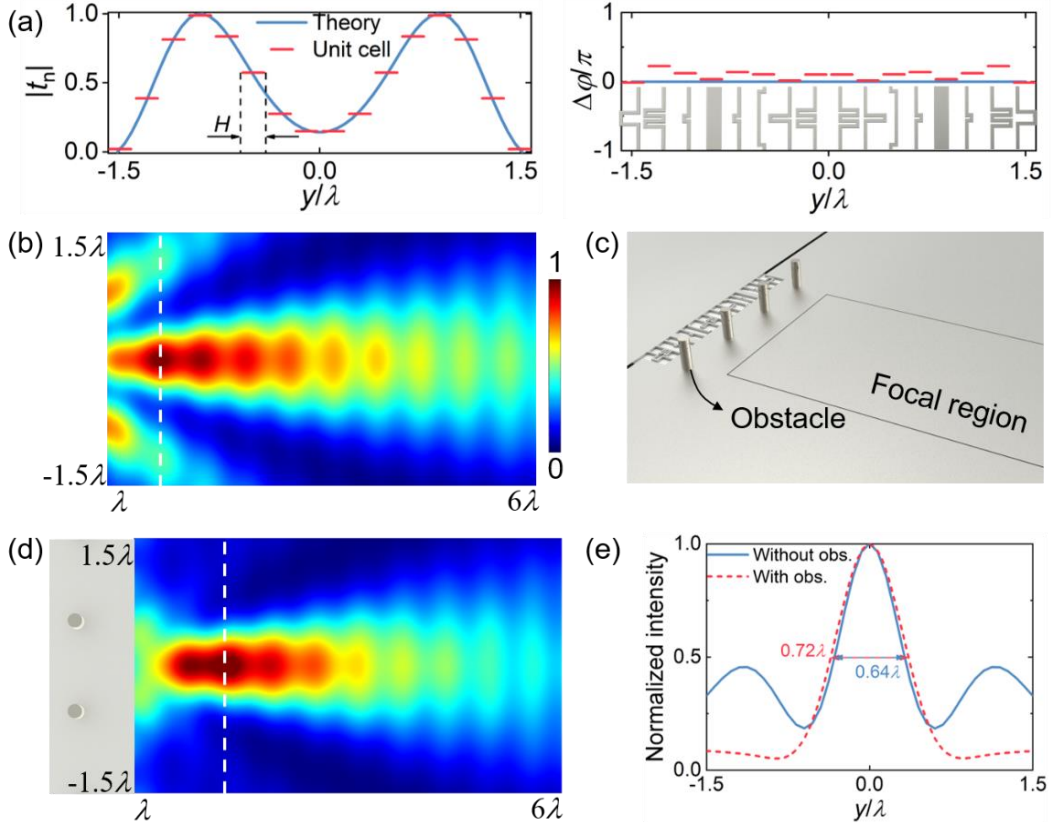

Fig. 8. (a) Profiles of transmission coefficient amplitudes $|t_n|$ and phase shifts $\Delta\varphi/\pi$ from theory and used unit cells for flexural wave focusing. (b) Normalized intensity distributions of the focal area from simulations. The white dashed line represents the cross section on focal spot along $y$ direction. (c) Illustration for the setup of four cylindrical obstacles. (d) Intensity distributions of the focal region after the scattering by obstacles. (e) Normalized intensity profiles along the white dashed lines in (b) and (d).

As shown in Fig. 8(c), we set four cylinders of radiuses 2.5 mm and heights 15 mm behind the metasurface as obstacles, with spacings 26.7 mm between neighboring pillars, to investigate the robustness against obstacles. Fig. 8(d) displays the intensity distributions of the focal region after the Airy-like beams are scattered by obstacles. A focal spot is also observed, which confirms the robust properties of present flexural wave focusing against obstacles. The comparison between the normalized intensity profiles on the cross sections of focal spots along $y$ direction is displayed in Fig. 8(e), for the cases with and without pillared obstacles. The full widths at half maximums are, respectively, $0.72\lambda$ and $0.64\lambda$ for the cases with and without obstacles, indicating a similar focusing characteristic. In addition, the side lobes are greatly suppressed by obstacles, and thus more energy is squeezed into the main lobe.

## 5. Experimental verification

To further validate the capability of present phase- and amplitude-shift elastic metasurfaces, experiment tests are carried out in this section for the metasurfaces of beam generation and flexural wave focusing. The experimental setup is shown in Fig. 9.

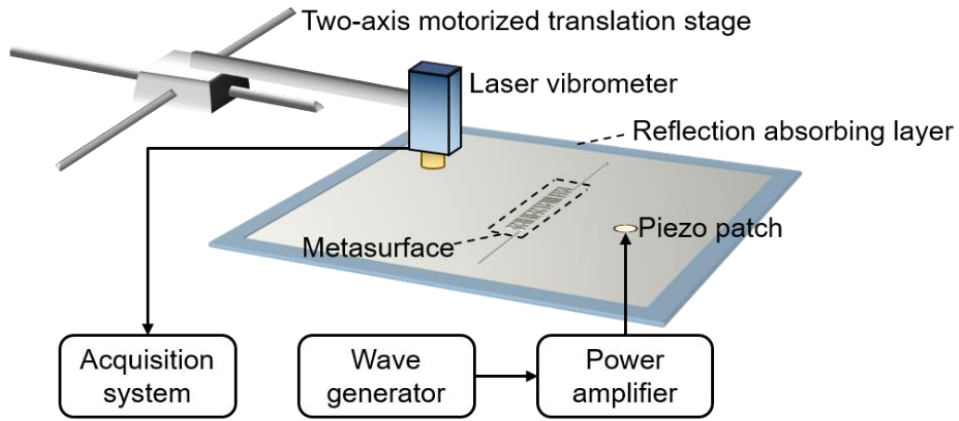

Fig. 9. Experimental setup for measuring the transmitted out-of-plane displacement fields.

A RIGOL DG4062 wave generator is used to exert a 3-cycle tone burst $F(t) = [1 - \cos(2\pi f_c t/3)] \sin(2\pi f_c t)$ with $f_c = 15$ kHz being the central frequency, which is magnified by a power amplifier to increase the signal-to-noise ratio. Piezoelectric patches driven by the power amplifier are bonded on fabricated plates as stimuluses. A Polytec NLV-2500 laser vibrometer, mounted on a two-axis motorized translation stage, is controlled to scan the transmitted out-of-plane displacement fields, which are stored onto computer via an oscilloscope of PicoScope 4000 series. To ensure precision of measurement, a layer of reflective film is pasted on the measurement area of fabricated plates. The metasurfaces of hollowed zigzag structure are synthesized by processing the 304-steel plates via metal wire cutting technique. The fabricated metasurfaces are shown in Fig. 10(a) and 11(a).

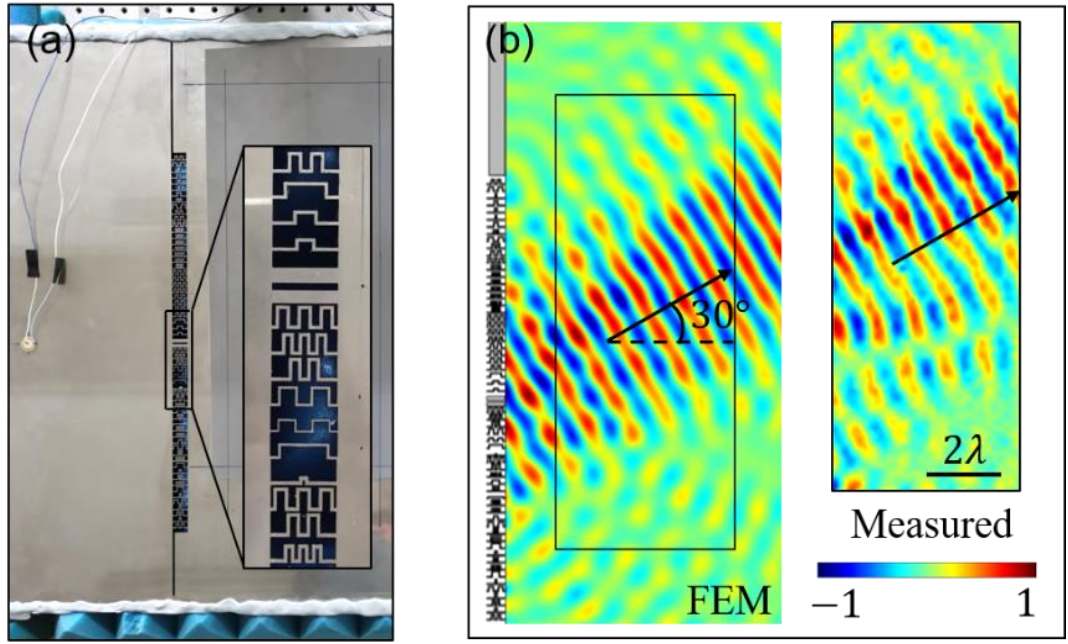

Fig. 10. Measured transmitted displacement field for the transformation from a cylindrical wave into a Gaussian beam with oblique angle 30°. (a) Fabricated plate with a hollow-structured metasurface and a piezoelectric patch glued as a point source. Inset is the enlarged view of a part of the metasurface. (b) Comparison between the transmitted displacement fields from FEM and experiment in frequency domain of 15 kHz.

Fig. 10(a) shows the partial view of fabricated plate ($800\times580\times1.5$ mm$^3$) for transforming a cylindrical wave into a Gaussian beam with oblique angle 30°. Inset is the partial enlargement of the hollow-structured metasurface. The measurement area covered by a layer of reflective film is of $380\times150$ mm$^2$. A circular piezoelectric patch of diameter 15 mm is bonded on the plate to excite a cylindrical wave. Blue-tack is attached along the boundaries of fabricated plates to cancel the effect of boundary reflection. Owing to the impedance mismatch, the blue-tack may not work perfectly on the 304-steel plates. To further eliminate boundary reflection, a time window is set to collect the portion of signal directly from the metasurface. The measurement is conducted at a spatial resolution of 5 mm ($\sim \lambda/6$) for working frequency 15 kHz, and the sampling frequency is taken to be 400 kHz. For the sake of comparison, the collected data in time domain are transformed into frequency domain via fast Fourier transform. Fig. 10(b) presents the displacement distributions in measurement area from FEM and experiment at operating frequency 15 kHz. A beam propagating along oblique angle 30° is seen in the measured wave field, coinciding with the FEM results. Differing from the FEM results, the amplitude distributions in a wavefront are not symmetric for the measured displacement field, which is attributed to the manufacturing errors.

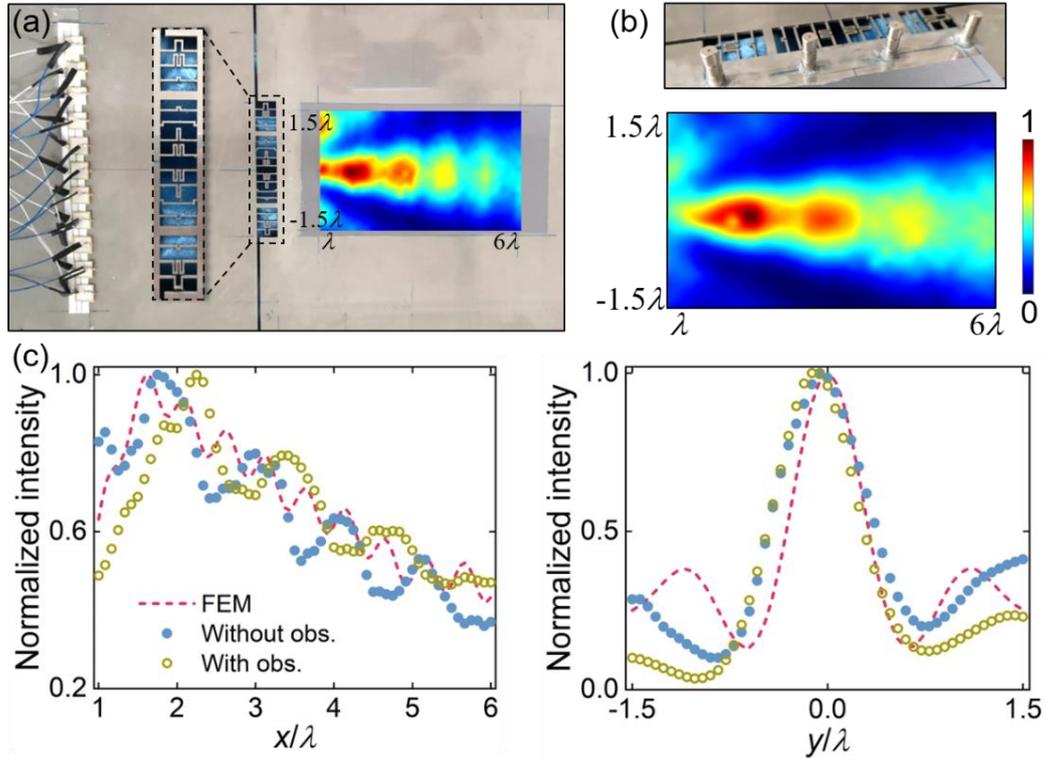

Fig. 11. Measured intensity patterns of focal area for the elastic focusing by only amplitude-shift modulations. (a) Measured intensity distribution of focal area, of which the background is the fabricated plate with a centered zigzag metasurface. Inset is the zoomed-in view of the metasurface. (b) Measured intensity distribution of focal area with additional pillared obstacles. Top panel illustrates experimental setup of obstacles. (c) Normalized intensity profiles on the cross sections of focal spots along $x$ and $y$ directions for FEM and experiment results. The dashed lines and symbols denote the FEM and experiment results, respectively.

Fig. 11(a) shows the measured intensity pattern of focal area for the flexural wave focusing by only amplitude modulations, of which the background is the synthesized plate with an array of 11 piezoelectric patches bonded to approximately mimic a normally incident Gaussian beam. The square piezoelectric patches ($13\times13$ mm$^2$), with a spacing of 5 mm between neighboring patches, are bonded to the fabricated plate with a distance of 150 mm from the metasurface and via a hard substrate stripe to uniformly distribute the amplitudes from individual sources. A good agreement is seen between the intensity patterns from FEM (shown in Fig. 8(b)) and experiment. Fig. 11(b) displays the measured intensity distribution with four pillared obstacles glued between the metasurface and focal area, which is illustrated by the top panel. The incident flexural wave is also concentrated in the focal area after scattering, and side lobes are suppressed by the obstacles, which are in coincidence with FEM results (Fig. 8(d)). The intensity distributions on side lobes in Figs. 11(a) and 11(b) are asymmetric, which is due to manufacturing errors. For the purpose of quantitative comparisons between the simulated and measured results, the normalized intensity profiles on the cross sections of focal spots along $x$ and $y$ directions are presented in Fig. 11(c). The simulated and experiment results are denoted by the dash lines and symbols, respectively. The measured focal length without obstacles, as well as the intensity trends in $x$ direction, are identical with FEM results. The focal length in experiment ($1.75\lambda$) is squeezed to be bigger by obstacles ($2.25\lambda$). From the intensity profiles in the $y$ direction, the measured full width at half maximum of main lobe ($0.76\lambda$) is bigger than that from FEM ($0.64\lambda$), which is attributed to that the Gaussian beam source in FEM has a more concentrated energy distribution than line sources in experiments. In addition, the side lobes suppressed by obstacles in experiments are clearly revealed by the intensity profiles along the $y$ direction.

## 6. Conclusions

In this work, we present a theoretical framework for Huygens-type elastic metasurfaces that can transform excitation fields into target patterns using artificial amplitude- and phase-shift interfaces, including a design rule and an intuitive prediction formula. The profiles of phase shifts and transmission coefficient amplitudes of metasurfaces are determined according to the discontinuities on interface between the incident and target fields. An intuitive prediction theory is developed by integrating the amplitude- and phase-shift into the classical Huygens-Fresnel principle. The generalized Snell's law can be well understood from the present theory by setting a unitary transmitted coefficient amplitude. We design two typical passive Huygens-type metasurfaces concerning flexural waves to demonstrate the functionalities of present theory: one is to transform a cylindrical wave into a Gaussian beam by phase and amplitude modulations, and the other one is to generate two symmetrical Airy-like beams to focus merely by amplitude modulations. We demonstrate the modulation capacity of zigzag unit cells in amplitude- and phase-shift and reveal the underlying mechanism. The transmission dips induced by twisting resonances can be adjusted by varying the defined spacings, which leads to the amplitude shifts decoupled from phase modulations. Theoretical calculations, numerical simulations and experimental tests agree well with each other for beam generation and flexural wave focusing. It is noted that present flexural wave focusing still works even though obstacles are placed behind the metasurface, manifesting its robustness.

The present work underlies the designs of elastic metasurfaces with phase- and amplitude-shift. The complete steering over elastic waves introduced by amplitude modulations may be of significance for diversified wave field transformations. Thus, the application prospect of elastic metasurfaces is hopefully broadened. For instance, the analog signal differentiating based on metasurfaces by amplitude modulations can be used for edge detections, as has been addressed in optics (Zhou et al., 2020), which has a promise for the detection of structural defects.

**Acknowledgements**

This work is supported by the NSFC (grant No. 11902239) and the young talent program of Shaanxi Association of Science and Technology (grant No. 20190501).

**Appendix A. Reduced GHFP to comprehend the generalized Snell's law**

We consider a flat Huygens-type metasurface in a planar Cartesian coordinate system, as shown in Fig. 12(a). An incident plane wave $W_1 = A_1 e^{ik(\cos\theta_{in} x + \sin\theta_{in} y)}$ and an linear phase shift $\Delta\varphi = k_m y$ along the metasurface at $x = 0$ are set here, where $\theta_{in}$ denotes the incident angle and $k_m$ is the gradient of phase shift. Based on the generalized Snell's law $k\sin\theta_t = k\sin\theta_{in} + k_m$, the transmitted field should be a plane wave with a propagating angle $\theta_t$ (Yu et al., 2011). Here we will show that the transmitted field can be calculated by GHFP in a more fundamental form of wave propagation. Base on Eq. (1), the terms in inclination factor are determined by $\cos(\boldsymbol{r}, \boldsymbol{n}) = \cos\theta_{in}$ and $\cos(\boldsymbol{s}, \boldsymbol{n}) = x/|\boldsymbol{s}|$ with $|\boldsymbol{s}| = \sqrt{x^2 + (y - y_m)^2}$. If we assume unitary transmission efficiency $|t_n| = 1$ of the metasurface, the transmitted field

should be

$$w = -\frac{i}{2\lambda} \int_{-B_u}^{B_u} A_1 \frac{e^{i(k\sin\theta_t y_m + k|s|)}}{\sqrt{|s|}} \left(\cos\theta_{in} + \frac{x}{|s|}\right) dy_m \qquad (A1)$$

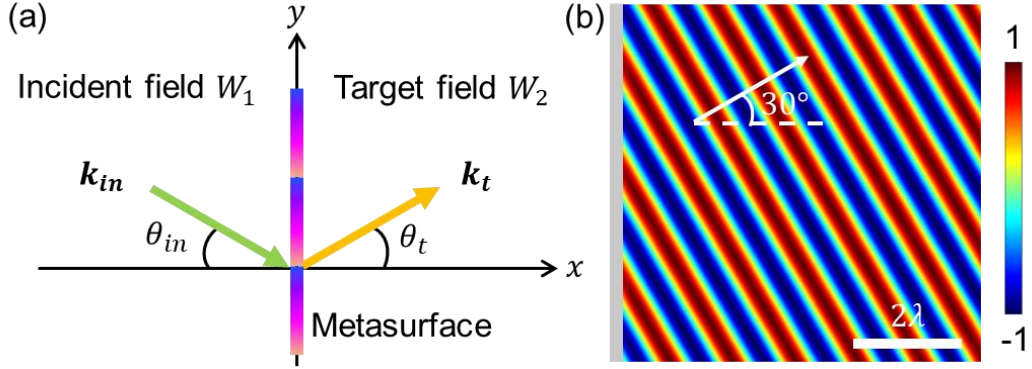

Fig. 12. Huygens-type metasurface with phase modulations to achieve anomalous transmission. (a) Diagram for a linear phase-shift metasurface redirecting an incident plane wave. (b) Transmitted wave field with $\theta_t = 30°$, which is calculated from Eq. (A1).

As an example, Fig. 12(b) displays the transmitted flexural wave field calculated from Eq. (A1), with incident angle $\theta_{in} = 0°$ and prescribed refraction angle $\theta_t = 30°$. The integral range $B_u$ is taken to be large enough to ensure the quality of the transmitted areas under observation. One can see ideal plane waves $W_2 = A_2 \exp(i\varphi_2)$ with $\varphi_2 = k\cos\theta_t\, x + k\sin\theta_t\, y$ propagating along prescribed directions, which verifies the geometric ray theory generalized Snell's law but in a wave theory.

**References**

Assouar, B., Liang, B., Wu, Y., Li, Y., Cheng, J.C., Jing, Y., 2018. Acoustic metasurfaces. Nat. Rev. Mater. 3, 460.

Born, M., Wolf, E., 2013. Principles of optics: electromagnetic theory of propagation, interference and diffraction of light. Elsevier.

Brule, S., Javalaud, E.H., Enoch, S., Guenneau., 2014. Experiments on seismic metamaterials: molding surface waves. Phys. Rev. Lett. 112, 133901.

Cao, L., Yang, Z., Xu, Y., Fan, S.W., Zhu, Y., Chen, Z., Li, Y., Assouar, B., 2020. Flexural wave

absorption by lossy gradient elastic metasurface. J. Mech. Phys. Solids. 143, 104052.

Chen, D.C., Zhu, X.F., Wu, D.J., Liu, X.J., 2019. Broadband Airy-like beams by coded acoustic metasurfaces. Appl. Phys. Lett. 114, 053504.

Chen, H.T., Taylor, A.J., Yu, N., 2016. A review of metasurfaces: physics and applications. Rep. Prog. Phys. 79, 076401.

Chen, J.S., Su, W.J., Cheng, Y., Li, W.C., Lin, C.Y., 2019. A metamaterial structure capable of wave attenuation and concurrent energy harvesting. J. Intell. Mater. Syst. Struct. 30, 2973.

Chen, J., Xiao, J., Lisevych, D., Shakouri, A., Fan, Z., 2018. Deep-subwavelength control of acoustic waves in an ultra-compact metasurface lens. Nat. Commun. 9, 1.

Chen, Y.Y., Huang, G.L., 2015. Active elastic metamaterials for subwavelength wave propagation control. Acta Mech. Sin. 31, 349.

Croxford, A.J., Wilcox, P.D., Drinkwater, B.W., Konstantinidis, G., 2007. Strategies for guided-wave structural health monitoring. Proc. R. Soc. A-Math. Phys. Eng. Sci. 463, 2961.

Estakhri, N.M., Alù, A., 2016. Wave-front Transformation with Gradient Metasurfaces. Phys. Rev. X. 6, 041008.

Goldsberry, B.M., Wallen, S.P., Haberman, M.R., 2019. Non-reciprocal wave propagation in mechanically-modulated continuous elastic metamaterials. J. Acoust. Soc. Am. 146, 782.

Jin, Y., Kumar, R., Poncelet, O., Mondain-Monval, O., Brunet, T., 2019. Flat acoustics with soft gradient-index metasurfaces. Nat. Commun. 10, 1.

Kim, M.S., Lee, W.R., Kim, Y.Y., Oh, J.H., 2018. Transmodal elastic metasurface for broad angle total mode conversion. Appl. Phys. Lett. 112, 241905.

Lee, H., Lee, J.K., Seung, H.M., Kim, Y.Y., 2018. Mass-stiffness substructuring of an elastic metasurface for full transmission beam steering. J. Mech. Phys. Solids. 112, 577.

Li, S., Xu, J., Tang, J., 2018. Tunable modulation of refracted lamb wave front facilitated by adaptive elastic metasurfaces. Appl. Phys. Lett. 112, 021903.

Li, Y., Assouar, M.B., 2016. Acoustic metasurface-based perfect absorber with deep subwavelength thickness. Appl. Phys. Lett. 108, 063502.

Li, Z., Palacios, E., Butun, S., Aydin, K., 2015. Visible-frequency metasurfaces for broadband anomalous reflection and high-efficiency spectrum splitting. Nano Lett. 15, 1615.

Liu, Y., Su, X., Sun, C.T., 2015. Broadband elastic metamaterial with single negativity by mimicking lattice systems. J. Mech. Phys. Solids. 74, 158.

Liu, Y., Liang, Z., Liu, F., Diba, O., Lamb, A., Li, J., 2017. Source illusion devices for flexural Lamb waves using elastic metasurfaces. Phys. Rev. Lett. 119, 034301.

Ma, G.C., Yang, M., Xiao, S.W., Yang, Z.Y., Sheng, P., 2014. Acoustic metasurface with hybrid resonances. Nat. Mater. 13, 873.

Mitra, M., Gopalakrishnan, S., 2016. Guided wave based structural health monitoring: A review. Smart Mater. Struct. 25, 053001.

Nassar, H., Chen, H., Norris, A.N., Haberman, M.R., Huang, G.L., 2017. Non-reciprocal wave propagation in modulated elastic metamaterials. Proc. R. Soc. A-Math. Phys. Eng. Sci. 473, 20170188.

Ni, X., Kildishev, A.V., Shalaev, V.M., 2013. Metasurface holograms for visible light. Nat. Commun. 4, 1.

Oh, J.H., Seung, H.M., Kim, Y.Y., 2017. Doubly negative isotropic elastic metamaterial for sub-wavelength focusing: Design and realization. J. Sound Vibr. 410, 169.

Park, J., Lee, D., Rho, J., 2020. Recent advances in non-traditional elastic wave manipulation by macroscopic artificial structures. Appl. Sci. 10, 547.

Pfeiffer, C., Grbic, A., 2013. Metamaterial Huygens' surfaces: tailoring wave fronts with reflectionless sheets. Phys. Rev. Lett. 110, 197401.

Semblat, J.F., Pecker, A., 2009. Waves and vibrations in soils: earthquakes, traffic, shocks, construction works. IUSS Press.

Serdyuk, V.M., Titovitsky, J.A., 2010. A simple analytic approximation for the refracted field at Gaussian beam incidence upon a boundary of absorbing medium. J. Electromagnet. Anal. Appl. 2, 640.

Seren, H.R., Keiser, G.R., Cao, L., Zhang, J., Strikwerda, A.C., Fan, K., Metcalfe, G.D., Wraback, M., Zhang, X., Averitt, R.D., 2014. Optically modulated multiband terahertz perfect absorber. Adv. Opt. Mater. 2, 1221.

Siviloglou, G.A., Broky, J., Dogariu, A., Christodoulides, D.N., 2007. Observation of accelerating Airy beams. Phys. Rev. Lett. 99, 213901.

Su, X., Lu, Z., Norris, A.N., 2018. Elastic metasurfaces for splitting SV- and P-waves in elastic solids. J. Appl. Phys. 123, 091701.

Sun, W.J., He, Q.O., Hao, J.M., Zhou, L., 2011. A transparent metamaterial to manipulate electromagnetic wave polarizations. Opt. Lett. 36, 927.

Tang, K., Qiu, C., Ke, M., Lu, J., Ye, Y., Liu, Z., 2014. Anomalous refraction of airborne sound through ultrathin metasurfaces. Sci. Rep. 4, 1.

Tian, Y., Wei, Q., Cheng, Y., Liu, X., 2017. Acoustic holography based on composite metasurface with decoupled modulation of phase and amplitude. Appl. Phys. Lett. 110, 191901.

Tufail, Y., Yoshihiro, A., Pati, S., Li, M.M., Tyler, W.J., 2011. Ultrasonic neuromodulation by brain stimulation with transcranial ultrasound. Nat. Protoc. 6, 1453.

West, P.R., Stewart, J.L., Kildishev, A.V., Shalaev, V.M., Shkunov, V.V., Strohkendl, F., Zakharenkov, R.K., Dodds, R.K., Byren, R., 2014. All-dielectric subwavelength metasurface focusing lens. Opt. Express. 22, 26212.

White, P., Clement, G., Hynynen, K., 2006. Longitudinal and shear mode ultrasound propagation in human skull bone. Ultrasound Med. Biol. 32, 1085.

Yoo, Y.J., Zheng, H.Y., Kim, Y.J., Rhee, J.Y., Kang, J.H., Kim, K.W., Cheong, H., Kim, Y.H., Lee, Y.P., 2014. Flexible and elastic metamaterial absorber for low frequency, based on small-size unit cell. Appl. Phys. Lett. 105, 041902.

Yu, N., Genevet, P., Kats, M.A., Aieta, F., Tetienne, J.P., Capasso, F., Gaburro, Z., 2011. Light propagation with phase discontinuities: generalized laws of reflection and refraction. Science. 334, 333.

Yu, N., Aieta, F., Genevet, P., Kats, M.A., Gaburro, Z., Capasso, F., 2012. A broadband, background-free quarter-wave plate based on plasmonic metasurfaces. Nano Lett. 12, 6328.

Zhang, J., Su, X., Liu, Y., Zhao, Y., Jing, Y., Hu, N., 2019. Metasurface constituted by thin composite beams to steer flexural waves in thin plates. Int. J. Solids. Struct. 162, 14.

Zheng, G., Mühlenbernd, H., Kenney, M., Li, G., Zentgraf, T., Zhang, S., 2015. Metasurface holograms reaching 80% efficiency. Nat. Nanotechnol. 10, 308.

Zhou, Y., Zheng, H., Kravchenko, I.I., Valentine, J., 2020. Flat optics for image differentiation. Nat. Photon. 14, 316.

Zhu, H., Semperlotti, F., 2016. Anomalous refraction of acoustic guided waves in solids with geometrically tapered metasurfaces. Phys. Rev. Lett. 117, 034302.

Zhu, R., Liu, X.N., Hu, G.K., Sun, C.T., Huang, G.L., 2014. Negative refraction of elastic waves at the deep-subwavelength scale in a single-phase metamaterial. Nat. Commun. 5, 1.